\documentclass[reprint, amssymb, amsmath, amsfonts, a4paper, aps, floatfix, showpacs]{revtex4-1}
\usepackage{tabularx, xspace, rotating, units, enumerate, ucs, upgreek, multirow}

\newcommand{\Epi}{\affiliation{Department of Epileptology, University of Bonn, Sigmund-Freud-Stra{\ss}e~25, 53105~Bonn, Germany}}
\newcommand{\HISKP}{\affiliation{Helmholtz Institute for Radiation and Nuclear Physics, University of Bonn, Nussallee~14--16, 53115~Bonn, Germany}}
\newcommand{\IZKS}{\affiliation {Interdisciplinary Center for Complex Systems, University of Bonn, Br\"uhler Stra\ss{}e~7, 53175~Bonn, Germany}}

\newcommand{\abs}[1]{\mathopen{}\mathclose\bgroup\left\lvert#1\aftergroup\egroup\right\rvert}
\newcommand{\norm}[1]{\mathopen{}\mathclose\bgroup\left\|#1\aftergroup\egroup\right\|}
\newcommand{\kl}[1]{\mathopen{}\mathclose\bgroup\left(#1\aftergroup\egroup\right)}
\newcommand{\klg}[1]{\mathopen{}\mathclose\bgroup\left\{#1\aftergroup\egroup\right\}}
\newcommand{\kle}[1]{\mathopen{}\mathclose\bgroup\left[#1\aftergroup\egroup\right]}
\newcommand{\kls}[1]{\mathopen{}\mathclose\bgroup\left\langle#1\aftergroup\egroup\right\rangle}

\newcommand{\defi}{\mathrel{\mathop:}=}

\newcommand{\zscore}{$z$\nobreakdash-score\xspace}

\newcommand{\textsubscript}[1]{\(_\text{#1}\)}

\newlength{\breite}

\begin{document}

\title{Surrogate-assisted analysis of weighted functional brain networks}

\author{Gerrit Ansmann}
\author{Klaus Lehnertz}
\Epi \HISKP \IZKS

\begin{abstract}
Graph-theoretical analyses of complex brain networks is a rapidly evolving field with a strong impact for neuroscientific and related clinical research.
Due to a number of confounding variables, however, a reliable and meaningful characterization of particularly functional brain networks is a major challenge.
Addressing this problem, we present an analysis approach for weighted networks that makes use of surrogate networks with preserved edge weights or vertex strengths.
We first investigate whether characteristics of weighted networks are influenced by trivial properties of the edge weights or vertex strengths (e.g., their standard deviations). If so, these influences are then effectively segregated with an appropriate surrogate normalization of the respective network characteristic.
We demonstrate this approach by re-examining, in a time-resolved manner, weighted functional brain networks of epilepsy patients and control subjects derived from simultaneous EEG/MEG recordings during different behavioral states.
We show that this surrogate-assisted analysis approach reveals complementary information about these networks, can aid with their interpretation, and thus can prevent deriving inappropriate conclusions.
\end{abstract}

\maketitle

\section{Introduction}
The human brain can be regarded as a complex network of interacting subsystems.
Over the past decade, network theory \cite{Strogatz2001, Albert2002, Newman2003, Boccaletti2006a, Arenas2008} has contributed significantly to our understanding of the normal and pathophysiological functioning of the brain \cite{Reijneveld2007, Bullmore2009, Stam2010a, Sporns2011a}.
In this approach, functional brain networks are usually derived from either direct or indirect measurements of neural activity.
Network vertices (or nodes) are associated with sensors that are placed such as to sufficiently capture the dynamics of different brain regions.
Network edges (or links) are associated with interactions between pairs of brain regions, which are assessed by evaluating some linear or nonlinear interdependence between their neural activities \cite{Pikovsky_Book2001, Pereda2005, Hlavackova2007, Lehnertz2009b}.
Functional brain networks have been shown to be neither entirely regular nor entirely random but to exhibit prominent topological properties, such as small mean shortest path lengths and a high level of clustering.
These and other network properties promise to allow an improved differentiation between various physiological and pathophysiological states and to be of relevance for clinical practice.
Recently, weighted networks, in which each edge is assigned a weight, have been shown to allow an improved description of functional networks underlying various brain pathologies such as schizophrenia \cite{Rubinov2009}, Alzheimer's disease \cite{Stam2009}, and epilepsy \cite{Ponten2009, Chavez2010, Horstmann2010}.

As for the analysis of binary networks, characteristics of weighted networks---such as the mean shortest path length and the clustering coefficient---are usually normalized using instances of an appropriate null model, or \textit{surrogates}, which can be obtained with Monte Carlo algorithms (for binary networks, see Refs.~\cite{Maslov2004, Sporns2004, Randrup2005, DelGenio2010}; for weighted networks, see Refs.~\cite{Barrat2004b, Serrano2008, Ansmann2011}).
There seems to be, however, no agreement on the use of surrogates for the investigation of weighted functional brain networks:
While some authors employ surrogates that preserve the edge weights \cite{Ponten2009, Stam2009,
Castellanos2011}, others use a degree-preserving approach \cite{Nakamura2009, Chavez2010, Dimitriadis2010} or no surrogates at all \cite{Horstmann2010, Zhang2011, Jin2011}.
It is conceivable that network characteristics are misinterpreted, e.g. due to a missing surrogate normalization or due to the use of inappropriate surrogates or normalizations.
Furthermore, results from most studies employing surrogates imply that characteristics of networks under analysis are close to those of the respective surrogates, which indicates that they may mainly reflect trivial properties of the data, that could have been assessed more easily without the network approach.

In order to prevent deriving inappropriate conclusions about weighted functional brain networks, we here present a surrogate-assisted analysis approach, in which possible influences of trivial properties on a given network characteristic are first identified and then segregated with an appropriate surrogate normalization.
We show that such a normalization can yield complementary information about such networks which could lead to an improved characterization of physiological and pathophysiological states of the brain.

\section{Materials and Methods}
To demonstrate the surrogate-assisted analysis approach, we here reanalyze weighted functional brain networks, which have been investigated by Ref.~\cite{Horstmann2010}.
In this study functional brain networks derived from simultaneously recorded electroencephalograms (EEG) and magnetoencephalograms (MEG) of epilepsy patients and of healthy controls have been analyzed.
Using different network construction rules, it was investigated whether the mean shortest path length $L$ and the clustering coefficient $C$ of weighted and binary functional brain networks differ between patients and controls and between different behavioral conditions (eyes open (EO) vs. eyes closed (EC)).
Although consistent differences could be observed---particularly with characteristics of weighted networks---the question remains, whether these findings indeed reflect topological properties of the investigated networks and not trivial properties of the data.

\subsection{Data acquisition and preprocessing}
In this section we give a brief overview of the acquisition and the preprocessing of this data, details of which can be found in the antecedent paper.
The study included 21 epilepsy patients and 23 healthy control subjects.
It was approved by the local Ethical Committees and all subjects gave their informed consent.
Of each subject EEG and MEG were recorded over a period of 30~minutes, during each half of which the subject had its eyes closed or opened, respectively, in a randomized order.
EEG data was recorded from 29 electrodes
with the right mastoid as physical reference.
MEG data was acquired with a 148 DC\nobreakdash-SQUID magnetometer whole head system, of whose 148 recording sites those on the lowermost ring were neglected to reduce influence of muscle artifacts, thus leaving 130 recording sites.
Both sets of data were recorded with \unit[254.31]{Hz} sampling rate, \unit[16]{bit} A/D~conversion and \unit[0\nobreakdash--50]{Hz} (EEG) or \unit[0.1\nobreakdash--50]{Hz} (MEG) bandwidth, respectively.
The influence of technical and physiological artifacts was reduced with a wavelet-based correction scheme.

To allow for a time-resolved network analysis (see, e.g., Refs.~\cite{Schindler2008a, Dimitriadis2010, Horstmann2010, Kuhnert2010, Bialonski2011b, Kramer2011, Schinkel2011}), the 30 minutes of recording were split into about 112 non-overlapping windows of \unit[16.1]{s} (4096 data points) each.
For statistical evaluation we followed the original paper and only regarded the windows 6 to 46 and 66 to 106 to avoid artifacts due to the change of conditions.

For each of these windows we extracted phases in a frequency-selective way with a wavelet transform using Morlet wavelets centered in the $\updelta$\nobreakdash- (\unit[0.5\nobreakdash--4]{Hz}), $\upvartheta$\nobreakdash- (\unit[4\nobreakdash--8]{Hz}), $\upalpha$\nobreakdash- (\unit[8\nobreakdash--13]{Hz}), $\upbeta$\textsubscript{1}\nobreakdash- (\unit[14\nobreakdash--20]{Hz}) or $\upbeta$\textsubscript{2}\nobreakdash-band (\unit[20\nobreakdash--30]{Hz}).
Phases also were extracted in a frequency-adaptive way using the Hilbert transform limited to the band \unit[0.5\nobreakdash--40]{Hz}.
From these phases the mean phase coherences $R_{ij} $ \cite{Mormann2000} between the signals from sensor $i$ and sensor $j$ were calculated.

In order to construct weighted functional brain networks, EEG and MEG recording sites were associated with network vertices.
Network weights were determined as:
\[W_{ij} \defi R_{ij} - \bar{R} + 1,\]
where $\bar{R}$ denotes the mean of all $R_{ij}$ with $i \neq j$ of a given network.
(These networks are denoted as WN\textsubscript{1} in Ref.~\cite{Horstmann2010}.)

\subsection{Network analysis}
\subsubsection{Network properties and characteristics}
The weights as defined above entirely describe a complete, weighted and undirected network, i.e., a weighted network with $W_{ij}=W_{ji}$ in which all possible edges exist.
We denote the number of vertices of such a network (here, the number of EEG or MEG recording sites) by $n$ and to simplify definitions we define the diagonal elements $W_{ii}$ to be zero.

The strength $S_i$ of vertex $i$ is defined as the sum of all adjacent weights: $S_i \defi \sum_{j=1}^n W_{ij}$.
In the following we regard the collection of all vertex strengths $\mathcal{S} \defi \klg{S_1, \ldots , S_n}$ and the collection of all edge weights $\mathcal{W} \defi \klg{W_{ij} \;\middle|\; 1 \leq i < j \leq n }$ (here, a collection or ``multiset'' is a set, in which members may appear multiple times).

As network characteristics we regard the clustering coefficient as suggested by Ref.~\cite{Onnela2005}, which was used in the study by Ref.~\cite{Horstmann2010} and has the advantage that its value is continuous for $W_{ij} \rightarrow 0$ \cite{Saramaki2007}:
\begin{equation}
\label{eqn:C}
C \defi \binom{n}{3}^{-1}
\sum\limits_{i=1}^n
\sum\limits_{j=1}^{i-1}
\sum\limits_{k=1}^{j-1}
\frac{\sqrt[3]{W_{ij} W_{jk} W_{ki}}}
{\max\kl{\mathcal{W}} }
\end{equation}
Since the maximum edge weight $\max\kl{\mathcal{W}}$ used for normalization may dominate this cluster coefficient (cf.~Fig.~\ref{fig:Hirnzeitreihen} and Ref.~\cite{Ansmann2011}), we also use the alternative definition:
\begin{equation}
K \defi \binom{n}{3}^{-1}
\sum\limits_{i=1}^n
\sum\limits_{j=1}^{i-1}
\sum\limits_{k=1}^{j-1}
\sqrt[3]{W_{ij} W_{jk} W_{ki}} = C \max\kl{\mathcal{W}}
\end{equation}

Note that no further normalization is necessary to eliminate the influence of the mean edge weight, since the latter has already been normalized to $1$ during network construction.

Furthermore we regard the mean shortest path length~$L$, for which we use the inverse of the weight of an edge as the length of that edge \cite{Newman2001c} and excluded the path from one vertex to itself from the mean:
\begin{equation}
L \defi
\binom{n}{2}^{-1}
\sum\limits_{i=1}^n
\sum\limits_{j=1}^{i-1}
\min_l
\min_{P\in\mathcal{P}_{ij}^l}
\sum\limits_{k=1}^{l-1}
W_{P_k P_{k+1}}^{-1}
,\end{equation}
where $\mathcal{P}_{ij}^l \defi \klg{P \in \klg{1,\ldots,n}^l \middle | P_1=i, P_l=j}$ is the set of all paths of binary length $l$ from $i$ to $j$, and $W_{ij}^{-1}=\infty$ if $W_{ij}=0$.

In the following we use $M$ as a placeholder for $C$, $K$, or $L$.

\subsubsection{Surrogate networks}

\label{surrogates}
In order to investigate whether the collection of all vertex strengths $\mathcal{S}$ or the collection of all edge weights $\mathcal{W}$ affect a network characteristic, we generate---for each functional brain network under investigation---4096 surrogates that preserve either $\mathcal{W}$ by shuffling the weights' positions \cite{Barrat2004b} or preserve $\mathcal{S}$ \cite{Ansmann2011}.
For a given functional brain network, we denote the collection of the strength-preserving surrogates by $\mathfrak{S}$ and the collection of the weight-preserving surrogates by $\mathfrak{W}$.
We consider network characteristics to be applied to these collections element-wise, e.g., $\overline{M\kl{\mathfrak{W}}}$ denotes the mean of $M$ over the weight-preserving surrogates of the network under consideration.
The decision on how many surrogates to generate usually involves a trade-off between computational effort and accuracy of sampling the distribution of a characteristic over the surrogates.
Using 4096 surrogates, we here concentrated on the latter.
Note, however, that fewer surrogates may be sufficient \cite{Schreiber2000a}.

\begin{figure}
	\includegraphics[width=\linewidth]{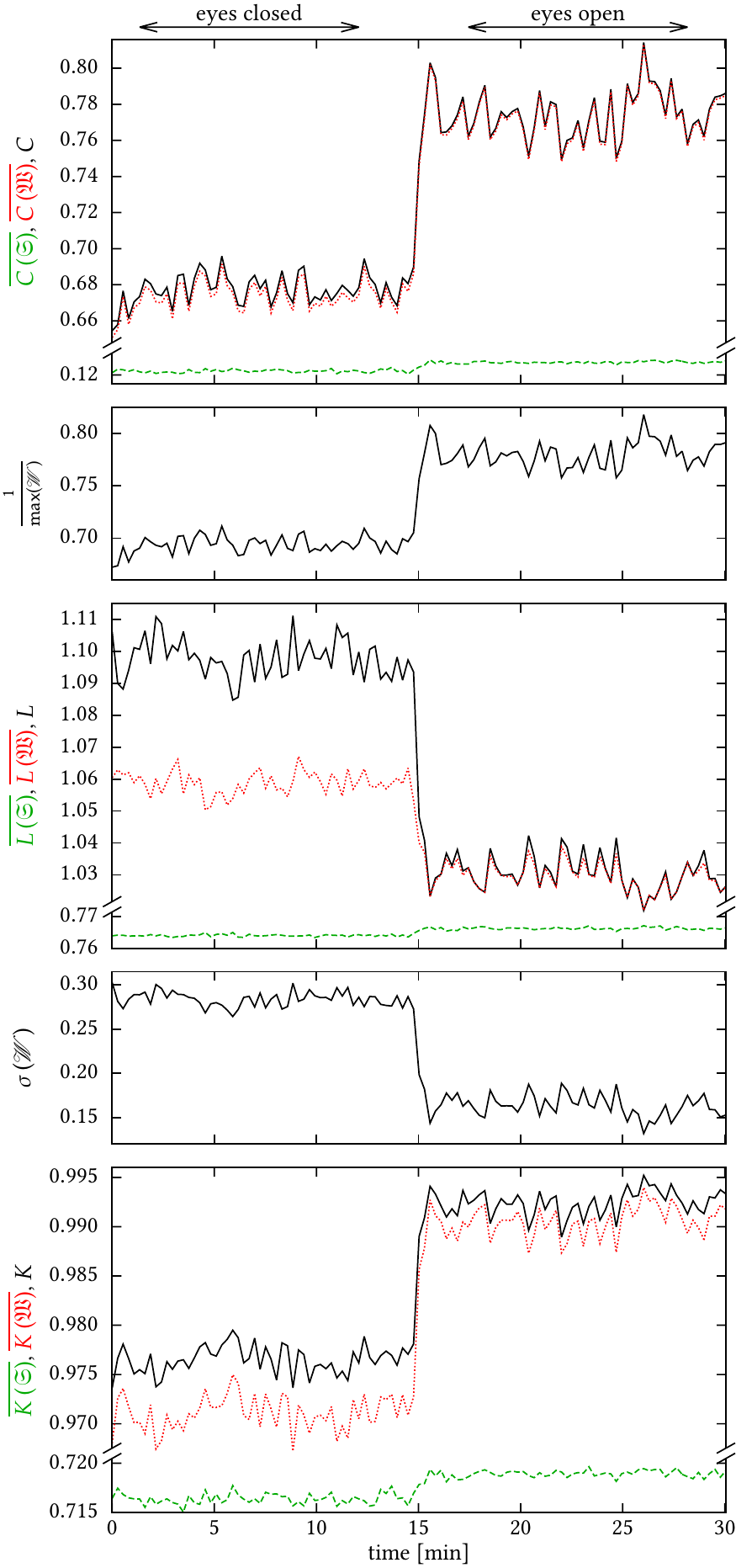}
	\caption{Exemplary temporal evolutions of some characteristics of the networks constructed from $\upvartheta$-band EEG data for one epilepsy patient as well as of the respective surrogates.
	Black, solid lines: original networks; green, dashed lines: mean over 4096 strength-preserving surrogates ($\mathfrak{S}$); red, dotted lines: mean over 4096 weight-preserving surrogates ($\mathfrak{W}$).
	Top row: The clustering coefficient $C$; third row: the mean shortest path $L$; bottom row: the alternate clustering coefficient $K$.
	The maxima of the standard deviation over the strength-preserving surrogates were about $0.02$ for each network characteristic.
	The maximum standard deviations over the weight-preserving surrogates were $0.0002$ for $C$, $0.001$ for $L$ and $0.0003$ for $K$.
	In the second and fourth row, we show the inverse of the maximum edge weight $\max\kl{\mathcal{W}}$ and the standard deviation of the edge weights, $\sigma\kl{\mathcal{W}}$, for comparison, both of the original networks only.
	The arrows above the plot indicate the intervals used for statistical evaluation.
	}
	\label{fig:Hirnzeitreihen}
\end{figure}

In Fig.~\ref{fig:Hirnzeitreihen} we show, as an example, the time courses of the clustering coefficients $C$ and $K$ as well as the mean shortest path length $L$ of the functional networks and of the corresponding surrogates for the data derived from an EEG recording.
The clustering coefficient $C$ is approximately equal for the original networks and for the weight-preserving surrogates.
The fact that this finding is less pronounced for $K$ together with the strong anti-correlation between $C$ and $\max\kl{\mathcal{W}}$ (Pearson's $r=-0.99$ for both behavioral conditions, EC and EO) confirm that $\max\kl{\mathcal{W}}$ has an overly strong influence on $C$ because of its use for normalization (cf.~Eqn.~\ref{eqn:C}).
For the mean shortest path length $L$, we also observe a strong correlation between the functional brain networks and their weight-preserving surrogates, however for EO only ($r=0.99$).
A property of the weight collection that we observe to be correlated to $L$ (EC: $r=0.93$; EO: $r=0.99$) as well as anti-correlated to $K$ (EC: $r=-0.95$; EO: $r=-0.98$) was the standard deviation of the edge weights $\sigma\kl{\mathcal{W}}$.
We observed weaker correlations between the original networks and the strength-preserving surrogates for $C$ (EC: $r=-0.21$; EO: $r=0.51$), $K$ (EC: $r=0.38$; EO: $r=0.49$), and $L$ (EC: $r=-0.23$; EO: $r=-0.54$).

Taking into account these preliminary results and our previous findings obtained from exemplary weighted functional brain networks \cite{Ansmann2011}, we conclude that the weight collection $\mathcal{W}$ has a considerable influence on the network characteristics $C$, $K$, and $L$.
To segregate this influence, we therefore use the weight-preserving surrogates $\mathfrak{W}$ for a normalization of these network characteristics in the following.

\subsubsection{Surrogate normalization of network characteristics}

From the many possible normalization schemes
we consider the following two:
First, we regard the relative difference of the characteristic between the functional brain networks and their surrogates (in the following abbreviated as RD): $\kl{M-\overline{M\kl{\mathfrak{W}}}}/\; \overline{M\kl{\mathfrak{W}}}$.
Second, we regard the difference of the characteristic between the functional brain networks and their surrogates in units of the standard deviation $\sigma$ of the characteristic over the surrogates (\zscore): $\kl{M-\overline{M\kl{\mathfrak{W}}}}/\sigma\kl{M\kl{\mathfrak{W}}}$.

Both approaches aim at quantifying, to which extent a characteristic of the network under consideration differs from what is to be expected from the network's weight collection~$\mathcal{W}$.
Since the \zscore also takes the range of this expectation into account, it may be considered to be more elaborate.
However, it requires $M\kl{\mathfrak{W}}$ to be approximately Gaussian and $\sigma\kl{M\kl{\mathfrak{W}}}>0$.
Of the network characteristics we analyzed~$L$ is most prone to failing to fulfill this requirement, since the weight collection~$\mathcal{W}$ may be such that the shortest paths in a surrogate network are most likely (or even certain) to be the direct ones, which results in $L\kl{\mathfrak{W}}$ to be heavily left-skewed (or to be delta-distributed).
We investigated the aforementioned requirement exemplarily for the collections $C\kl{\mathfrak{W}}$, $K\kl{\mathfrak{W}}$, and $L\kl{\mathfrak{W}}$ the time course of whose means are shown in Fig.~\ref{fig:Hirnzeitreihen}:
The $C\kl{\mathfrak{W}}$ and $K\kl{\mathfrak{W}}$ had absolute skewnesses of $0.47\pm0.04$ and absolute (excess) kurtoses of $0.31\pm0.14$.
Thus we consider them not to deviate too far from a normal distribution for the purposes of a $z$-score.
The $L\kl{\mathfrak{W}}$ had absolute skewnesses of $0.26\pm0.35$ and absolute kurtoses of $0.22\pm0.52$, however, only after neglecting some strong outliers and collections with $\sigma\kl{L\kl{\mathfrak{W}}}=0$, possible reasons for which have been given above.
Thus the \zscore of $L$ has to be handled with care.

Note that for both surrogate normalization schemes the contribution of $\max\kl{\mathcal{W}}$ is eliminated (alongside with that of~$\mathcal{W}$) and thus the surrogate normalizations of~$C$ are identical to the respective ones of~$K$.

\subsection{Statistical analyses}
\label{statistics}
For the statistical analyses we regard the subject-wise means over each condition of the following characteristics:

\begin{itemize}
	\item the RD and the \zscore of the clustering coefficient $C$ and of the mean shortest path length $L$ to investigate the complementary information yielded by these surrogate normalizations;
	\item $C$ and $L$ as reference for comparison with the original study \cite{Horstmann2010}.
	\item $K$, $\overline{C\kl{\mathfrak{W}}}$, $\overline{L\kl{\mathfrak{W}}}$, $\sigma\kl{\mathcal{W}}$, and $\frac{1}{\max\kl{\mathcal{W}}}$ to investigate, to which extent the results of the original study could have been obtained with more simple properties of the data.
\end{itemize}

We applied the Mann-Whitney-Wilcoxon test \cite{Mann1947} to identify possible differences of these network properties and characteristics between patients and controls\footnote{The requirement that the tested distributions are identical except for an offset was tested insofar that Levene's test \cite{Levene1960} did reject equality of variances with a significance level of $0.05$ in 16 of 264 cases of the comparisons.} and the Wilcoxon signed-rank test \cite{Wilcoxon1945} to identify possible differences of these network properties and characteristics between behavioral states.

To account for multiple testing (24~tests: patient vs. control group for EO and EC as well as EO vs. EC for both groups, each for 6~frequency bands), we applied the procedure by Ref.~\cite{Benjamini1995} to control the false discovery rate (FDR) at 0.05.
Note, however, that the individual tests might be correlated, e.g., one would to some extent expect the results for the EO condition to be similar to those for the EC condition.
Therefore the FDR control may be overly conservative and its results are to be taken as an estimate only.
Because of this and to obtain comparability to Ref.~\cite{Horstmann2010}, we report results after FDR control separately.

For all tests we used a significance level of $0.05$.

\section{Results}

Before presenting in detail differences of the clustering coefficients $C$ and $K$, the mean shortest path length $L$, and their related characteristics between patients and controls and between behavioral states, we checked whether our findings from section~\ref{surrogates} can be generalized to the entire data:
The mean correlation coefficients (calculated for all subjects, behavioral conditions, frequency bands, and recording techniques) between the original networks and weight-preserving surrogates amounted to $r=0.9996\pm0.0005$ for $C$, $r=0.91\pm0.10$ for $K$, and $r=0.91\pm0.18$ for $L$.

\subsection{Clustering coefficient and related properties}
\setlength{\tabcolsep}{6pt}
\begin{table}
	\begin{tabular}{@{} c | c | c | c | cccccc}
		& & comp. & data & bb & $\updelta$ & $\upvartheta$ & $\upalpha$ & $\upbeta_1$ & $\upbeta_2$\\\hline\hline
		\multirow{8}{*}{$C$}  & \multirow{4}{*}{\begin{sideways}EEG\hspace*{0.5em}\end{sideways}}  & EC & PG & & & $\downarrow$& & & \\\cline{4-4}
 & & vs. EO & CG & & $\downarrow$& $\Downarrow$& & & $\uparrow$\\\cline{3-10}
 & & PG & EC & $\uparrow$& $\uparrow$& & & $\uparrow$& \\\cline{4-4}
 & & vs. CG & EO & $\uparrow$& $\uparrow$& & & $\uparrow$& $\uparrow$\\\cline{2-10}
 & \multirow{4}{*}{\begin{sideways}MEG\hspace*{0.5em}\end{sideways}}  & EC & PG & & & & & & \\\cline{4-4}
 & & vs. EO & CG & & $\downarrow$& & & & \\\cline{3-10}
 & & PG & EC & & & & & & \\\cline{4-4}
 & & vs. CG & EO & & & & & & \\\hline\hline		\multirow{8}{*}{$\overline{C\kl{\mathfrak{W}}}$}  & \multirow{4}{*}{\begin{sideways}EEG\hspace*{0.5em}\end{sideways}}  & EC & PG & & & $\downarrow$& & & \\\cline{4-4}
 & & vs. EO & CG & & $\downarrow$& $\Downarrow$& & & $\uparrow$\\\cline{3-10}
 & & PG & EC & & $\uparrow$& & & $\uparrow$& \\\cline{4-4}
 & & vs. CG & EO & $\uparrow$& $\uparrow$& & & $\uparrow$& $\uparrow$\\\cline{2-10}
 & \multirow{4}{*}{\begin{sideways}MEG\hspace*{0.5em}\end{sideways}}  & EC & PG & & $\downarrow$& & & & \\\cline{4-4}
 & & vs. EO & CG & & $\downarrow$& & & & \\\cline{3-10}
 & & PG & EC & & & & & & \\\cline{4-4}
 & & vs. CG & EO & & & & & & \\\hline\hline		\multirow{8}{*}{$\frac{1}{\max\kl{\mathcal{W}}}$}  & \multirow{4}{*}{\begin{sideways}EEG\hspace*{0.5em}\end{sideways}}  & EC & PG & & & $\Downarrow$& & & \\\cline{4-4}
 & & vs. EO & CG & & $\downarrow$& $\Downarrow$& & $\uparrow$& $\uparrow$\\\cline{3-10}
 & & PG & EC & $\uparrow$& $\Uparrow$& & & $\uparrow$& \\\cline{4-4}
 & & vs. CG & EO & $\Uparrow$& $\Uparrow$& & & $\Uparrow$& $\uparrow$\\\cline{2-10}
 & \multirow{4}{*}{\begin{sideways}MEG\hspace*{0.5em}\end{sideways}}  & EC & PG & & $\downarrow$& & & & \\\cline{4-4}
 & & vs. EO & CG & & $\downarrow$& & & & \\\cline{3-10}
 & & PG & EC & & & & & & \\\cline{4-4}
 & & vs. CG & EO & & & & & & \\\hline\hline		\multirow{8}{*}{$K$}  & \multirow{4}{*}{\begin{sideways}EEG\hspace*{0.5em}\end{sideways}}  & EC & PG & $\Downarrow$& & $\Downarrow$& $\Downarrow$& $\Downarrow$& \\\cline{4-4}
 & & vs. EO & CG & $\Downarrow$& & $\Downarrow$& $\Downarrow$& $\Downarrow$& \\\cline{3-10}
 & & PG & EC & & & & & & \\\cline{4-4}
 & & vs. CG & EO & & & & & & \\\cline{2-10}
 & \multirow{4}{*}{\begin{sideways}MEG\hspace*{0.5em}\end{sideways}}  & EC & PG & & & & & & \\\cline{4-4}
 & & vs. EO & CG & $\downarrow$& & & & & \\\cline{3-10}
 & & PG & EC & $\downarrow$& $\downarrow$& & & & \\\cline{4-4}
 & & vs. CG & EO & $\Downarrow$& $\Downarrow$& & & & \\\hline\hline		\multirow{8}{*}{\settowidth{\breite}{$\tfrac{C-\overline{C\kl{\mathfrak{W}}}}{\sigma\kl{C\kl{\mathfrak{W}}}}$}\begin{minipage}[c]{\breite}\[\tfrac{C-\overline{C\kl{\mathfrak{W}}}}{\sigma\kl{C\kl{\mathfrak{W}}}}\]\[=\]\[\tfrac{K-\overline{K\kl{\mathfrak{W}}}}{\sigma\kl{K\kl{\mathfrak{W}}}}\]\end{minipage}}  & \multirow{4}{*}{\begin{sideways}EEG\hspace*{0.5em}\end{sideways}}  & EC & PG & $\Downarrow$& & $\Downarrow$& $\downarrow$& $\Downarrow$& $\Downarrow$\\\cline{4-4}
 & & vs. EO & CG & $\Downarrow$& & $\downarrow$& $\Downarrow$& $\Downarrow$& $\Downarrow$\\\cline{3-10}
 & & PG & EC & & & & & $\uparrow$& \\\cline{4-4}
 & & vs. CG & EO & & & & & & \\\cline{2-10}
 & \multirow{4}{*}{\begin{sideways}MEG\hspace*{0.5em}\end{sideways}}  & EC & PG & & & & & & \\\cline{4-4}
 & & vs. EO & CG & & & & & & \\\cline{3-10}
 & & PG & EC & & & & & & \\\cline{4-4}
 & & vs. CG & EO & & & & & & \\\hline\hline		\multirow{8}{*}{\settowidth{\breite}{$\tfrac{C-\overline{C\kl{\mathfrak{W}}}}{\overline{C\kl{\mathfrak{W}}}}$}\begin{minipage}[c]{\breite}\[\tfrac{C-\overline{C\kl{\mathfrak{W}}}}{\overline{C\kl{\mathfrak{W}}}}\]\[=\]\[\tfrac{K-\overline{K\kl{\mathfrak{W}}}}{\overline{K\kl{\mathfrak{W}}}}\]\end{minipage}}  & \multirow{4}{*}{\begin{sideways}EEG\hspace*{0.5em}\end{sideways}}  & EC & PG & & & & & & $\Downarrow$\\\cline{4-4}
 & & vs. EO & CG & $\downarrow$& & & & $\downarrow$& $\Downarrow$\\\cline{3-10}
 & & PG & EC & & $\uparrow$& $\uparrow$& & & \\\cline{4-4}
 & & vs. CG & EO & & $\Uparrow$& & & & \\\cline{2-10}
 & \multirow{4}{*}{\begin{sideways}MEG\hspace*{0.5em}\end{sideways}}  & EC & PG & & & & & & \\\cline{4-4}
 & & vs. EO & CG & & & & & & \\\cline{3-10}
 & & PG & EC & & & & & & \\\cline{4-4}
 & & vs. CG & EO & & & & & & \\\hline\hline	\end{tabular}
	\caption{Summary of findings obtained for comparing the clustering coefficients $C$ and $K$ as well as related characteristics between (1) the eyes-closed (EC) and the eyes-opened (EO) condition (findings are reported separately for the patient (PG) and the control group (CG)) and (2) between the patient and the control group (findings are reported separately for the two conditions).
	Characteristics for weighted networks constructed from EEG and MEG recordings in different frequency bands (bb: broadband signals).
	Upward/downward arrows indicate that the values of network characteristics during EC or for PG were significantly larger/smaller.
	Double arrows indicate significant differences after FDR control for multiple testing.}
	\label{tab:CTabelle}
\end{table}

In Table~\ref{tab:CTabelle} we list statistically significant differences of the clustering coefficients $C$ and $K$ and of related characteristics between the patient and the control group and between the eyes-closed and the eyes-open condition.
In general, findings were almost identical for the clustering coefficient $C$ of the original functional brain networks and of the weight-preserving surrogates ($\overline{C\kl{\mathfrak{W}}}$) as well as for the inverse of the maximum edge weight ($\frac{1}{\max\kl{\mathcal{W}}}$).
From this, together with the observations for the exemplary time series (see Sect.~\ref{surrogates}) and with findings from Ref.~\cite{Ansmann2011}, we conclude that for the networks under investigation, $C$ mainly reflected properties of the weight collection $\mathcal{W}$, mostly $\max\kl{\mathcal{W}}$.
The findings for $K$ differed from those for $C$ and displayed several statistically significant differences (even after FDR control) between the behavioral states but were mostly matching those for the standard deviation of the edge weights $\sigma\kl{\mathcal{W}}$ (see Table~\ref{tab:LTabelle})---just having the opposite sign.
This confirms our results for the exemplary time series for which $K$ and $\sigma\kl{\mathcal{W}}$ were anti-correlated.

We observed the \zscore{}s of $C$ and $K$ of functional brain networks constructed from EEG (but not MEG) data to be significantly smaller during the EC state (see also upper half of Fig.~\ref{fig:Hists}).
This finding holds for both groups and all frequency bands, except the $\updelta$\nobreakdash-band.
For the RD of $C$, we observed fewer significant differences between behavioral states.
Of both surrogate normalizations we observed only few significant differences between the patient group and the control group (see also upper half of Fig.~\ref{fig:Hists}).

\subsection{Mean shortest path length and related properties}

\begin{table}
	\begin{tabular}{@{} c | c | c | c | cccccc}
		& & comp. & data & bb & $\updelta$ & $\upvartheta$ & $\upalpha$ & $\upbeta_1$ & $\upbeta_2$\\\hline\hline
		\multirow{8}{*}{$L$}  & \multirow{4}{*}{\begin{sideways}EEG\hspace*{0.5em}\end{sideways}}  & EC & PG & $\Uparrow$& & $\uparrow$& $\Uparrow$& $\Uparrow$& \\\cline{4-4}
 & & vs. EO & CG & $\Uparrow$& & $\Uparrow$& $\Uparrow$& $\Uparrow$& \\\cline{3-10}
 & & PG & EC & & $\Uparrow$& & & & \\\cline{4-4}
 & & vs. CG & EO & & $\Uparrow$& & & & \\\cline{2-10}
 & \multirow{4}{*}{\begin{sideways}MEG\hspace*{0.5em}\end{sideways}}  & EC & PG & & & & $\uparrow$& & \\\cline{4-4}
 & & vs. EO & CG & & & & $\uparrow$& & \\\cline{3-10}
 & & PG & EC & & $\uparrow$& & & & \\\cline{4-4}
 & & vs. CG & EO & & $\uparrow$& & & & \\\hline\hline		\multirow{8}{*}{$\overline{L\kl{\mathfrak{W}}}$}  & \multirow{4}{*}{\begin{sideways}EEG\hspace*{0.5em}\end{sideways}}  & EC & PG & $\uparrow$& & $\Uparrow$& $\Uparrow$& $\Uparrow$& \\\cline{4-4}
 & & vs. EO & CG & $\uparrow$& & $\Uparrow$& $\Uparrow$& $\Uparrow$& \\\cline{3-10}
 & & PG & EC & & & & & & \\\cline{4-4}
 & & vs. CG & EO & & & & & & \\\cline{2-10}
 & \multirow{4}{*}{\begin{sideways}MEG\hspace*{0.5em}\end{sideways}}  & EC & PG & & & & & & \\\cline{4-4}
 & & vs. EO & CG & & & & & & \\\cline{3-10}
 & & PG & EC & & $\uparrow$& & & & \\\cline{4-4}
 & & vs. CG & EO & & $\uparrow$& & & & \\\hline\hline		\multirow{8}{*}{$\sigma\kl{\mathcal{W}}$}  & \multirow{4}{*}{\begin{sideways}EEG\hspace*{0.5em}\end{sideways}}  & EC & PG & $\uparrow$& & $\Uparrow$& $\Uparrow$& $\Uparrow$& \\\cline{4-4}
 & & vs. EO & CG & $\uparrow$& & $\Uparrow$& $\Uparrow$& $\Uparrow$& \\\cline{3-10}
 & & PG & EC & & & & & & \\\cline{4-4}
 & & vs. CG & EO & & & & & & \\\cline{2-10}
 & \multirow{4}{*}{\begin{sideways}MEG\hspace*{0.5em}\end{sideways}}  & EC & PG & & & & & $\uparrow$& \\\cline{4-4}
 & & vs. EO & CG & $\uparrow$& & & & & \\\cline{3-10}
 & & PG & EC & $\uparrow$& $\uparrow$& & & & \\\cline{4-4}
 & & vs. CG & EO & $\uparrow$& $\uparrow$& & & & \\\hline\hline		\multirow{8}{*}{$\tfrac{L-\overline{L\kl{\mathfrak{W}}}}{\sigma\kl{L\kl{\mathfrak{W}}}}$}  & \multirow{4}{*}{\begin{sideways}EEG\hspace*{0.5em}\end{sideways}}  & EC & PG & $\uparrow$& & & $\Uparrow$& $\Uparrow$& \\\cline{4-4}
 & & vs. EO & CG & $\Uparrow$& & $\Uparrow$& $\Uparrow$& $\Uparrow$& \\\cline{3-10}
 & & PG & EC & $\Uparrow$& $\Uparrow$& $\Uparrow$& & & \\\cline{4-4}
 & & vs. CG & EO & $\Uparrow$& $\Uparrow$& $\Uparrow$& & & $\Downarrow$\\\cline{2-10}
 & \multirow{4}{*}{\begin{sideways}MEG\hspace*{0.5em}\end{sideways}}  & EC & PG & & & $\uparrow$& $\uparrow$& & \\\cline{4-4}
 & & vs. EO & CG & & & & $\uparrow$& $\uparrow$& \\\cline{3-10}
 & & PG & EC & & $\uparrow$& $\uparrow$& & $\uparrow$& \\\cline{4-4}
 & & vs. CG & EO & & $\uparrow$& & & $\uparrow$& $\uparrow$\\\hline\hline		\multirow{8}{*}{$\tfrac{L-\overline{L\kl{\mathfrak{W}}}}{\overline{L\kl{\mathfrak{W}}}}$}  & \multirow{4}{*}{\begin{sideways}EEG\hspace*{0.5em}\end{sideways}}  & EC & PG & $\uparrow$& & & $\Uparrow$& $\Uparrow$& \\\cline{4-4}
 & & vs. EO & CG & $\Uparrow$& & $\Uparrow$& $\Uparrow$& $\Uparrow$& \\\cline{3-10}
 & & PG & EC & $\Uparrow$& $\Uparrow$& $\uparrow$& & & \\\cline{4-4}
 & & vs. CG & EO & $\Uparrow$& $\Uparrow$& $\Uparrow$& & & \\\cline{2-10}
 & \multirow{4}{*}{\begin{sideways}MEG\hspace*{0.5em}\end{sideways}}  & EC & PG & & & $\uparrow$& $\uparrow$& & \\\cline{4-4}
 & & vs. EO & CG & & & & $\uparrow$& $\uparrow$& \\\cline{3-10}
 & & PG & EC & & $\uparrow$& $\uparrow$& & $\uparrow$& \\\cline{4-4}
 & & vs. CG & EO & & $\uparrow$& & & $\uparrow$& $\uparrow$\\\hline\hline	\end{tabular}
	\caption{Same as Tab.~\ref{tab:CTabelle} but for the mean shortest path length $L$ and related characteristics.}
	\label{tab:LTabelle}
\end{table}

Table~\ref{tab:LTabelle} summarizes our findings for the mean shortest path length and related characteristics.
We observed almost the same statistically significant differences either between behavioral states or between patients and controls for $L$ of the original functional brain networks and of the weight-preserving surrogates ($\overline{L\kl{\mathfrak{W}}}$) as well as for the standard deviation of the edge weights ($\sigma\kl{\mathcal{W}}$).
As for the clustering coefficient $C$ we thus derive that the mean shortest path length $L$ of the networks under investigation mainly reflected properties of the weight collection $\mathcal{W}$, mostly $\sigma\kl{\mathcal{W}}$.

\begin{figure*}
	\includegraphics[width=\linewidth]{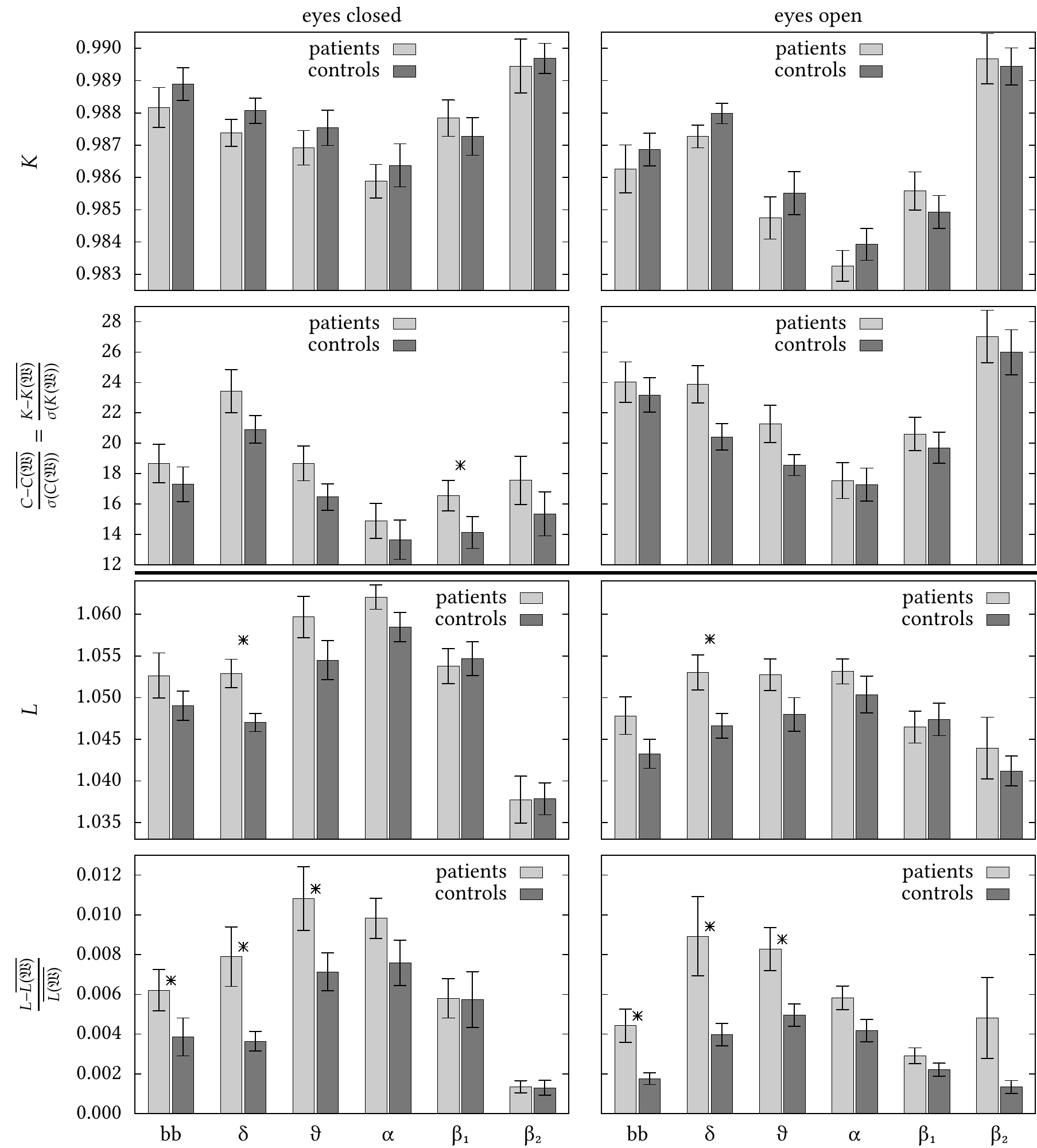}
	\caption{Group means and standard errors of exemplary characteristics of weighted functional brain networks derived from EEG data in different frequency bands (bb: broadband signals) for the eyes-closed (left column) and the eyes-open condition (right column). Upper half: the clustering coefficient $K$ and its $z$-score (with respect to the weight-preserving surrogates) .
	Lower half: the mean shortest path length $L$ and its relative difference to the mean of $L$ over the weight-preserving surrogates ($\overline{L\kl{\mathfrak{W}}}$).
	Stars indicate significant differences ($p<0.05$; no FDR control) between the patient and the control group.}
	\label{fig:Hists}
\end{figure*}

For the functional brain networks derived from either EEG or MEG data, the findings for the RD were almost identical to those for the \zscore:
For both groups, the surrogate-normalized mean shortest path length was significantly larger during EC for EEG-derived networks in the $\upalpha$\nobreakdash- and the neighboring frequency bands.
EEG-derived networks of the patient group exhibited a significantly larger surrogate-normalized mean shortest path length for data from the
$\updelta$\nobreakdash- and $\upvartheta$\nobreakdash-bands, independent of the behavioral state (see also lower half of Fig.~\ref{fig:Hists}).
For the MEG-derived networks we also observed the surrogate-normalized mean shortest path length to exhibit higher values for the patient group or for the eyes-closed condition, respectively. These differences were, however, less pronounced and extended to the $\upbeta$\nobreakdash-bands.

The fact that significant differences in the $\updelta$-band between patients and controls for the EEG data could also be observed for $L$ but not for $\overline{L\kl{\mathfrak{W}}}$ indicates that the differences observed for $L$ were not caused by properties of the weight collection and were enhanced by the surrogate normalization (see also lower half of Fig.~\ref{fig:Hists}).

\section{Discussion and Conclusions}
Network theory provides a powerful framework for studying the complex functioning of the brain.
Despite its rapidly increasing success, interpretation of empirically estimated network characteristics remains a challenging issue.
Do values of some network characteristic reveal properties of the investigated functional brain network or do they merely reflect trivial properties of the data that may be accessible more easily, thus questioning the network approach as an overly complicated description of simple aspects of the data?
To address this question, surrogate normalizations have been applied in the past, however, little focus has been put on the choice of surrogates and normalization schemes.

We have described a surrogate-assisted analysis approach that helps to detect possible influences of properties of either the weight or the strength collection on characteristics of complete weighted networks and, if appropriate, to effectively segregate these influences with appropriate surrogate normalizations (see also Ref.~\cite{Andrzejak2011} and references therein).
This analysis approach makes use of surrogate networks with a preserved weight \cite{Barrat2004b} or strength collection \cite{Ansmann2011}, which can be generated with low computational effort.
It can provide complementary information about network characteristics and can aid with their interpretation, and thus it can prevent deriving inappropriate conclusions.

We demonstrated the usefulness of the approach by exemplarily re-examining the clustering coefficient and the mean shortest path length of weighted functional brain networks derived from simultaneous EEG and MEG recordings of epilepsy patients and healthy controls during different behavioral states \cite{Horstmann2010}.
For these networks we identified the maximum or, respectively, the standard deviation of the weight collection to dominate the network characteristics.
It is conceivable that other network characteristics \cite{Costa2007, Rubinov2010} are as well influenced by trivial properties of the data.
With commonly used normalization schemes (relative difference of the network characteristic between the functional brain networks and the surrogates as well as the \zscore) we could segregate these trivial influences.
This way, we observed differences, that can be traced back---with high confidence---to nontrivial properties of the underlying functional brain networks.
The differences across groups were consistent across states and the differences between behavioral states were consistent across groups.
These findings underline the high value of the network approach to further improve our understanding of the complex functioning of the brain.

\section{Acknowledgments}
We are grateful to Stephan Bialonski, Christian Geier, Marie-Therese Kuhnert, and Alexander Rothkegel for helpful comments.
This work was supported by the Deutsche Forschungsgemeinschaft (Grant No.~LE660/4\nobreakdash-2).

\nocite{Levene1960}

\end{document}